\documentclass{article}

\usepackage{graphicx}
\usepackage{amsmath}
\usepackage{amssymb}
\usepackage{siunitx}
\usepackage{authblk}

\begin{document}

\title{Compact Multi-Spectral Pushframe Camera for Nano-Satellites}

\author[1]{Yoann Noblet} 
\author[2]{Stuart Bennett}
\author[1]{Paul F. Griffin}
\author[2]{Paul Murray}
\author[2]{Stephen Marshall}
\author[3]{Wojciech Roga}
\author[1]{John Jeffers}
\author[1]{Daniel Oi}

\affil[1]{Department of Physics, SUPA, University of Strathclyde, Glasgow, G4 0NG, UK}
\affil[2]{Department of Electronic and Electrical Engineering, University of Strathclyde, Glasgow G1 1XQ, UK}
\affil[3]{National Institute of Information and Communications Technology, Koganei, Tokyo 184-8795, Japan}


\maketitle



\begin{abstract}
In this paper we present an evolution of the single-pixel camera architecture, called 'pushframe', which addresses the limitations of pushbroom cameras in space-based applications. In particular, it is well-suited to observing fast moving scenes while retaining high spatial resolution and sensitivity. We show that the system is capable of producing colour images with good fidelity and scalable resolution performance. The principle of our design places no restriction on the spectral range to be captured, making it suitable for wide infrared imaging.
\end{abstract}

\section{Introduction}
The single pixel camera (SPC) was first demonstrated in 2008 \cite{duarte2008single}, offering a simpler and cheaper alternative to conventional 2D arrays outside of the visible band region. This is especially true in the infrared region where commercial 2D sensors are either unavailable or very expensive. The second motivation behind SPCs was to implement direct compressive sensing (CS) of the imaged scene \cite{candes2005decoding,candes2006near}. Compressive sensing relies on the fact that natural scenes are highly compressible in different bases such as wavelets, i.e. they have a sparse representation \cite{starck2010sparse}. Sampling in an incoherent basis allows reconstruction with a number of measurements related to the sparsity of the underlying signal, which may be orders of magnitude smaller than the overall image size.

SPCs gained interest in the last decade and have found applications in infrared and terahertz imaging \cite{edgar2015simultaneous,stantchev2016noninvasive}, three-dimensional imaging \cite{sun2016single}, ghost imaging \cite{shapiro2008computational,sun2012normalized}, non-line-of-sight three-dimensional imaging \cite{musarra2019non}, gas leak detection \cite{gibson2017real}, and biomedical imaging and microscopy \cite{lochocki2016single,tajahuerce2014image,radwell2014single}. The compact nature and ease of operation of SPCs made them perfect candidates for medical applications when the time of exposure for radiation and size of the imager need to be minimized.

SPCs require multiple exposures of the scene to form an image (4096 measurements in order to obtain a resolution of 64 x 64 pixels), and as such the resolution of the final image is fixed by the spatial light modulator, such as a DMD (digital micromirror device) array, and not the detection element's size. Therefore, by using different types of DMD patterns, highly flexible data acquisition modes can be used as required by the imaging situation. Masks can be adapted to provide high resolution in only the regions of interest, similar to the foveal structure of the retina \cite{phillips2017adaptive}. Alternatively, the global resolution of the DMD pattern can be reduced without the need for interpolation or resampling after acquisition. Observation using scene-adapted masks has several advantages in remote sensing. In coastal monitoring for example, details from deeper waters or inland are not of interest. Masks can be designed to reject these regions and only provide high resolution information in the coastal areas. This greatly reduces the amount of data or sensor readings needed to be acquired in the first place to reconstruct the regions of interest as well as cutting down onboard processing or data reduction postcapture. Masks can also be used to block off bright parts of the scene, reducing glare in darker areas of interest and increasing the dynamic range \cite{qiao2015design}.

For multispectral imaging applications, single pixel techniques enable cost-efficient imaging at wavelengths where high-efficiency, low-noise detector arrays are not readily or cheaply available. Furthermore, because DMDs are inherently broadband compared with lenses, they allow for compact, shared-aperture multispectral optical configurations \cite{edgar2015simultaneous}. Additionally, all spectral bands are automatically registered, obviating the requirement for re-alignment. Together with a spectral separator, e.g. a diffraction grating or dichroic mirrors, only a single photodetector in each imaged wavelength is needed, instead of an array. A camera based on a single pixel multispectral imager would be able to image simultaneously in UV/visible, near-IR, and mid-IR bands, using commercially available DMDs and simple detectors.

However, one of the main limitations of SPCs is the relatively long acquisition time required to capture the image, during which the scene must not change. There is a trade-off between resolution and speed which renders these imagers confined to applications where either time or resolution is not an issue. More recently other techniques based on pushbroom imaging have been developed, which work like a line-scan camera and allow fast, high resolution imaging \cite{fowler2014compressive,arablouei2016fast}. Pushbroom cameras do not rely on a single detector but rather on a linear array which essentially decreases the number of measurements needed to recover the scene without decreasing the resolution \cite{arnob2018compressed}, compared to an SPC. This technique, however, implies a scanning motion perpendicular to the aperture slit. This renders the technique sub-optimal when observing a static scene, but can be overcome by scanning the aperture slit across the scene \cite{abdo2018dual}. Similarly to SPCs, pushbroom cameras also offer the possibility of hyperspectral imaging by placing a dispersive element and a spatial light modulator to project the wanted spectral component on to the linear detector, leading to the reconstruction of a datacube, i.e. two spatial dimensions and one spectral dimension \cite{fowler2014compressive,arablouei2016fast,arnob2018compressed}. Hyperspectral imaging has found applications in agriculture and forestry \cite{adao2017hyperspectral}, medical imaging \cite{lu2014medical}, food monitoring \cite{gowen2007hyperspectral,polak2019use}, and remote sensing \cite{van2012multi}.

In this paper we present a novel SPC imaging technique working in the so-called pushframe mode which addresses the limitations of pushbroom cameras in space-based applications. In particular, the rapid motion of a satellite across the Earth restricts the exposure time and consequently has a significant impact on the signal to noise ratio of the acquired image. In a pushbroom camera the exposure time is usually limited by image smear caused by the Earth moving across one ground sample distance. Indeed, in prevalent SPC models the patterns of light illuminating an object are changing quickly enough for the object to be regarded as unchanged, so the dynamics of the scene determine the typical exposure time per pattern. As the imaging spatial resolution increases with the number of patterns, for a given capture period higher spatial resolution requires reducing the exposure time per pattern, which leads to degradation of imaging quality or infeasibility.

\section{Working principle}
\subsection{Pushframe Camera}
\begin{figure}
	\centering\includegraphics[width=\textwidth]{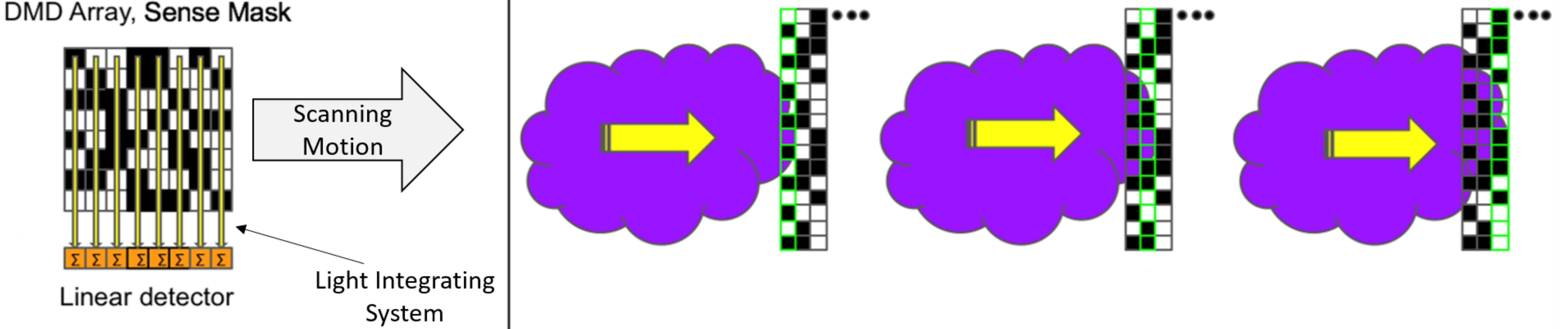}
	\caption{Schematic representation of a pushframe imaging camera. Light from a given column of a scene is reflected from different columns of the DMD as the scene moves across the pattern. The total light reflected from each column is measured by a pixel from the linear detector using integrating optics such as a set of cylindrical lenses. The signals from each element of the linear detector are gathered repeatedly, each time the scene moves across one pattern column, which allows us to reconstruct each given column of the scene.}
	\label{fig:pushframe}
\end{figure}

The novel imaging technique presented in this paper relies on the scanning motion of the camera across the scene or, conversely, the motion of the scene across the camera field of view, to apply different mask patterns to each vertical strip of the scene. This technique alleviates the need of displaying many different patterns as proposed in \cite{henriksson2017imaging,baraniuk2006method}, thus allowing moving objects to be detected. Unlike the pushframe camera introduced in \cite{anderson2009challenges}, which works like a snapshot camera using the camera's motion to capture different 'strips' of the scene, the pushframe concept introduced in this paper allows for compressive sensing to be used \cite{rani2018systematic,kyriakides2016target} and does not inherit the limitations of a snapshot architecture, i.e. image smear, limited multi-spectral abilities and large data sets. Our novel imaging concept is illustrated in Figure~\ref{fig:pushframe} where by displaying a Hadamard pattern, i.e. a single two dimensional mask that is constructed from a full matrix of linearly independent one dimensional patterns, a moving scene would cross each column of the matrix. The array of photodetectors would then capture the light intensity summed from each column and that process is repeated each time the scene moves one pattern column. The 1D optical integration as well as multiple exposures of each ground cell provide an increased signal-to-noise ratio (SNR) compared to other imaging methods. This could find applications in low-light imaging, especially in the infrared region where the quantum efficiency of the sensors still lags behind that of the more mature and affordable visible technology.

The rows and columns of the Hadamard matrix form a linearly independent set of binary numbers that can be used to construct a set of linearly independent masks to be displayed on the DMD. In this way each measurement provides a new piece of information about the image and ensures an efficient way of reconstructing the scene.

Though we can use a fixed mask pattern, the use of a DMD allows adaptive measurement schemes \cite{deutsch2009adaptive}. Instead of the static pattern used in Figure~\ref{fig:pushframe}, consider a situation where the measurements from the first few columns can be used to adapt the pattern of the following columns that the object will encounter later. This flexibility is advantageous for compressive sensing, machine learning, object-tracking, shape recognition and related techniques of scene analysis without image recording. This versatility is not present in conventional pushbroom imaging. Our pushframe technique is an intermediate solution between a pushbroom and a traditional snapshot camera, which inherits advantages of both techniques, such as hyperspectral capabilities and high signal to noise ratio, which is a crucial advantage for remote sensing.

\subsection{Experimental Setup} \label{Setup}
Our experimental setup is based on a commercial DMD (Vialux V-7000 module). This module uses a DLP7000 DMD from Texas Instruments with a 14.0\,$\times$\,10.5~mm micromirror array and a resolution of 1024 by 768 pixels. This gives a mirror pitch of 13.7~$\mu$m, making the device compatible with short-wave infrared (SWIR) imaging. Each micromirror can be individually tilted at $\pm$\ang{12} along the diagonal axis, making it a bistable spatial light modulator. The dimensions of the module are quite modest; the controller electronics fit on a 71\,$\times$\,68~mm circuit board, and the DMD board is 67\,$\times$\,50~mm. The patterns can be loaded on to the on-board memory of 32~Gb and the device has a maximum switching rate of 22.7~kHz. We load a Hadamard pattern on to the DMD where the scene will be imaged. A Hadamard matrix is square, with the number of rows and columns being a power of two.  Assuming the pattern is scaled only by integer factors, we are therefore limited to using an area of 512 by 512 pixels on the DMD.

\begin{figure}
	\centering\includegraphics[width=7.5cm]{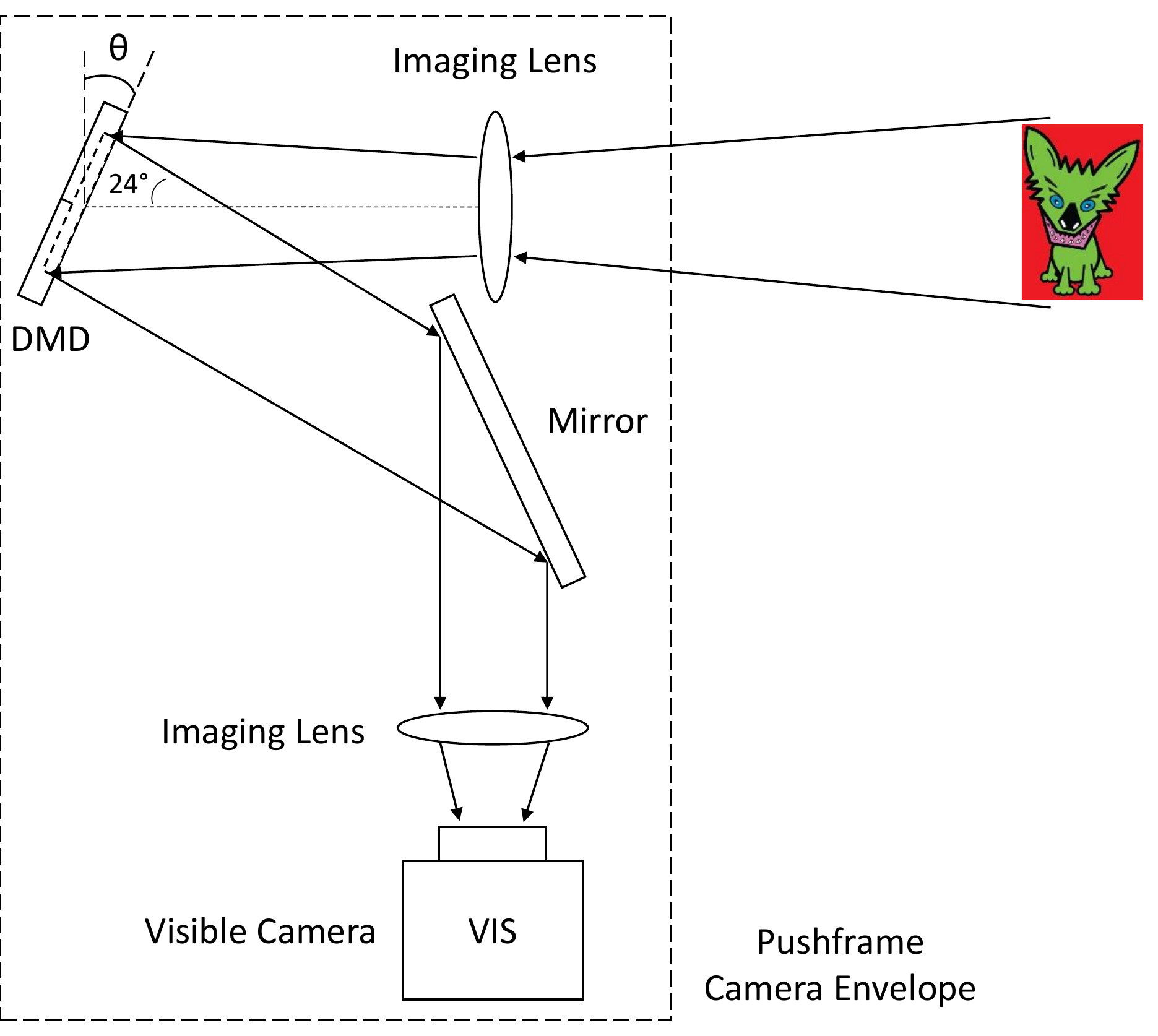}
	\caption{Experimental setup. The fore-optics consist of a doublet lens focussing the image of the scene on to the DMD. The scene intensity is modulated by the pattern on the DMD, reflecting part of the light towards the detection arm through the after-optics. The angle $\theta$ between the optical axis and the DMD image plane is set at \ang{24}. A 2D sensor is used to simulate the effect of a 1D optical integrator via post-processing.}
	\label{fig:setup}
\end{figure}

As shown in Figure~\ref{fig:setup}, the scene (Dog) is imaged on to the DMD using an achromatic doublet lens (Thorlabs AC254-075-A). However, unlike a planar mirror, the DMD's individually tilting micromirrors mean different parts of the reflected scene have different optical path lengths, and so can only simultaneously be in sharp focus at the camera if the camera is skewed: in turn causing a poor focus on the DMD pattern. When experimenting we obtained better results by having a sharp pattern and blurry scene on the camera ($\theta$=\ang{24}, i.e. double the tilt angle of the micromirrors, with the pattern displayed on the DMD being parallel to the detector plane) rather than a blurry pattern and a sharp scene ($\theta$=\ang{0}, i.e. the DMD plane being perpendicular to the imaging axis). For ease of alignment, because the square micromirrors tilt along their diagonal, we rotated the DMD by \ang{45} such that the mirrors tilt on a plane parallel to the imaging axis. This means that the scene has to move along the same axis, i.e. diagonally at \ang{45}. The image of the scene is then reflected off a mirror and re-imaged on the camera (Thorlabs DCC3240M) using another achromatic doublet lens (Thorlabs AC254-125-A). 
	
The light integration system shown in Figure~\ref{fig:pushframe} happens here computationally, i.e. we sum the intensity of all the pixels forming each row in software, simulating a linear detector. We chose to work with a conventional 2D array as it offers the advantage of being able to observe the pre-summation image and evaluate whether it suffers from optical aberrations, skew, or vignetting. This gives us the opportunity to foresee any imaging problems that might occur when using the 1D integrating optics, as once the image has been compressed down in one dimension, it becomes very difficult to determine what in particular is affecting a reconstructed image's quality. Preliminary work on the 1D integrating optics has shown that it is possible to achieve the required compression, within a nano-satellite envelope, using a set of custom-made toroidal lenses. The image of the Hadamard pattern on the camera needs to exhibit high contrast as well as good sharpness as it is essential to avoid pixel crosstalk.
	
The scene used in our pushframe camera is displayed on an LCD monitor. This allows us to control the step size of the moving scene very accurately, moving the image consistently by one pixel on the monitor between each capture, over the total required movement of the scene. The LCD monitor also has the advantage of being backlit with good uniformity which is essential in a pushframe camera (see Section \ref{Reconstruction}). In practice, for space observation, the movement of the satellite in orbit will be known precisely and the uniformity of the scene would be ideal.
	
In order to set up the camera, a Hadamard pattern image occupying 512\,$\times$\,512 pixels is loaded on the DMD board and displayed as a static pattern. The DMD pattern is then imaged on the camera until maximum sharpness is obtained. The next step is to display a scene on the monitor and adjust the distance between the scene and the first imaging lens until a sharp image of the scene is achieved on the camera. Finally we need to match the displacement of the scene with the size of one Hadamard pattern pixel on the DMD. It is only possible to move the scene by a minimum of one pixel on the monitor which means that the distance between the monitor and the first lens needs to be further adjusted until the displacement on the screen corresponds to the size of one row of pixels on the DMD pattern. To increase the contrast, it is imperative to minimise stray light reaching both the DMD and the camera, which means that the imaging system was fully enclosed. The relatively small footprint of the experimental setup (60\,$\times$\,50~cm) allows for easy enclosure of the whole pushframe camera system which also makes it a good candidate for the small payloads associated with nano-satellites.

	\begin{figure}
	\centering\includegraphics[width=8cm]{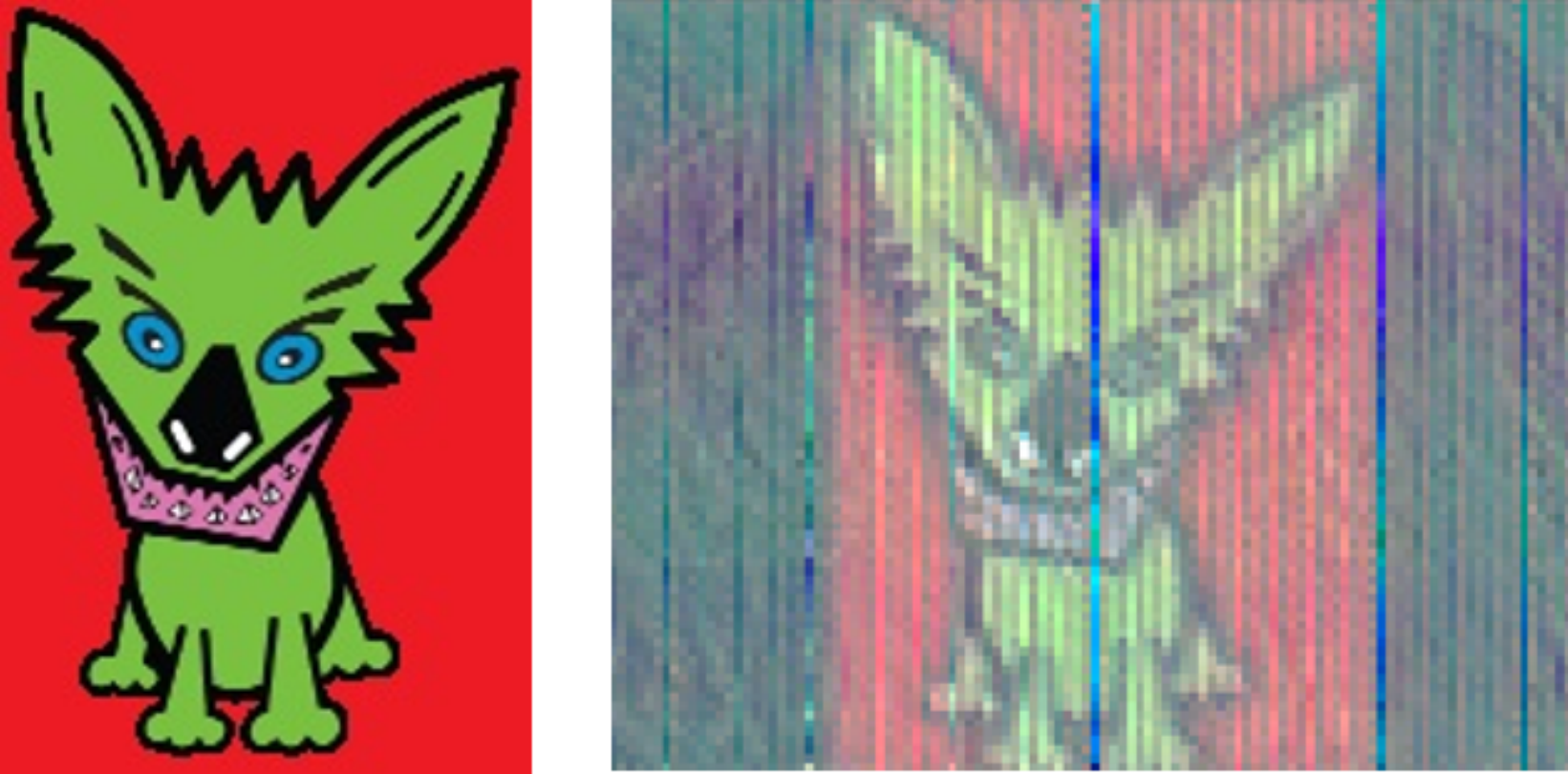}
	\caption{(Left) Image projected by the LCD monitor that is then moved across the DMD pattern. (Right) Reconstructed image of the object depicted on the left hand side.}
	\label{fig:64}
\end{figure}

\section{Reconstruction Algorithm} \label{Reconstruction}
With a Hadamard sampling pattern, reconstruction is	straightforward. Each column of the reconstructed image is formed by adding together weighted versions of the columns of the Hadamard sampling pattern. The weights are the sum of the light intensity sampled by the corresponding pattern column, while that column was 
masking the relevant column of the external scene.

Therefore, if we index the columns of the DMD pattern with variable $i$, and the scene moves across the DMD from column 0 to column $n$, with the column of the Hadamard matrix at column $i$ being the vector $H_i$, and the sum of column $i$'s samples being $S_i$, a column first imaged at time step $t=0$ ($C\rvert_{_{t=0}}$) is reconstructed by
\begin{equation}
C\rvert_{_{t=0}} = H_0S_0\rvert_{_{t=0}} + 
H_1S_1\rvert_{_{t=1}} + \dots + H_nS_n\rvert_{_{t=n}}\text{.}
\label{eq:recon}
\end{equation}

In ideal conditions this would be sufficient, but the weightings ($S_i$) must be correct for an accurate reconstruction. In practice, the weightings, $S_i$, are degraded by stray light, sensor noise and non-uniform illumination (due either to optical aberrations or imperfections in the LCD's backlight). Consider the reconstruction of a uniform scene: each Hadamard column should contribute equally, but if the illumination of the DMD is twice as bright at the column $n$ side compared to that at column 0, $S_n$ will be twice the value of $S_0$ and the final vector $C$ will erroneously consist of a lot more of $H_n$ than $H_0$.

The non-ideal contributions can be partially compensated for during reconstruction, prior to applying Equation \ref{eq:recon}, via a \emph{per column} flat-field correction. In using column sums for this correction, only information that would be available from an optical integration implementation is needed. Retaining compatibility with an optical implementation does have a downside however, that illumination variation within the summed columns cannot be calibrated out: instead it will degrade the reconstruction. Thus it is essential that the scene lighting and optical path have very high uniformity.

\section{Experimental Results}
\subsection{Preliminary Results}
Early results obtained with this system were acquired using a 128\,$\times$\,128 pixels Hadamard pattern displayed by 512\,$\times$\,512 pixels on the DMD array. The DMD array was oriented such that its plane was parallel to the object plane, i.e. (with reference to Figure~\ref{fig:setup}, $\theta$=\ang{0}. The low resolution allows for better signal-to-noise ratio as well as limiting the pixel crosstalk to a minimum. It also makes matching the moving scene's step-size with the column-width of the Hadamard pattern easier to achieve. The result can be seen on the right side of Figure~\ref{fig:64}. The capture sequence starts when the to-be-imaged object enters the first row of the Hadamard pattern, triggering the camera to capture the corresponding frame. The second frame is captured when the object moves to the second row of the Hadamard pattern and so on until the object has completely moved across and out of the pattern. The reconstruction of the image seen in Figure~\ref{fig:64} is then achieved by using the algorithm introduced in Section \ref{Reconstruction}.

\begin{figure}
	\centering\includegraphics[width=8cm]{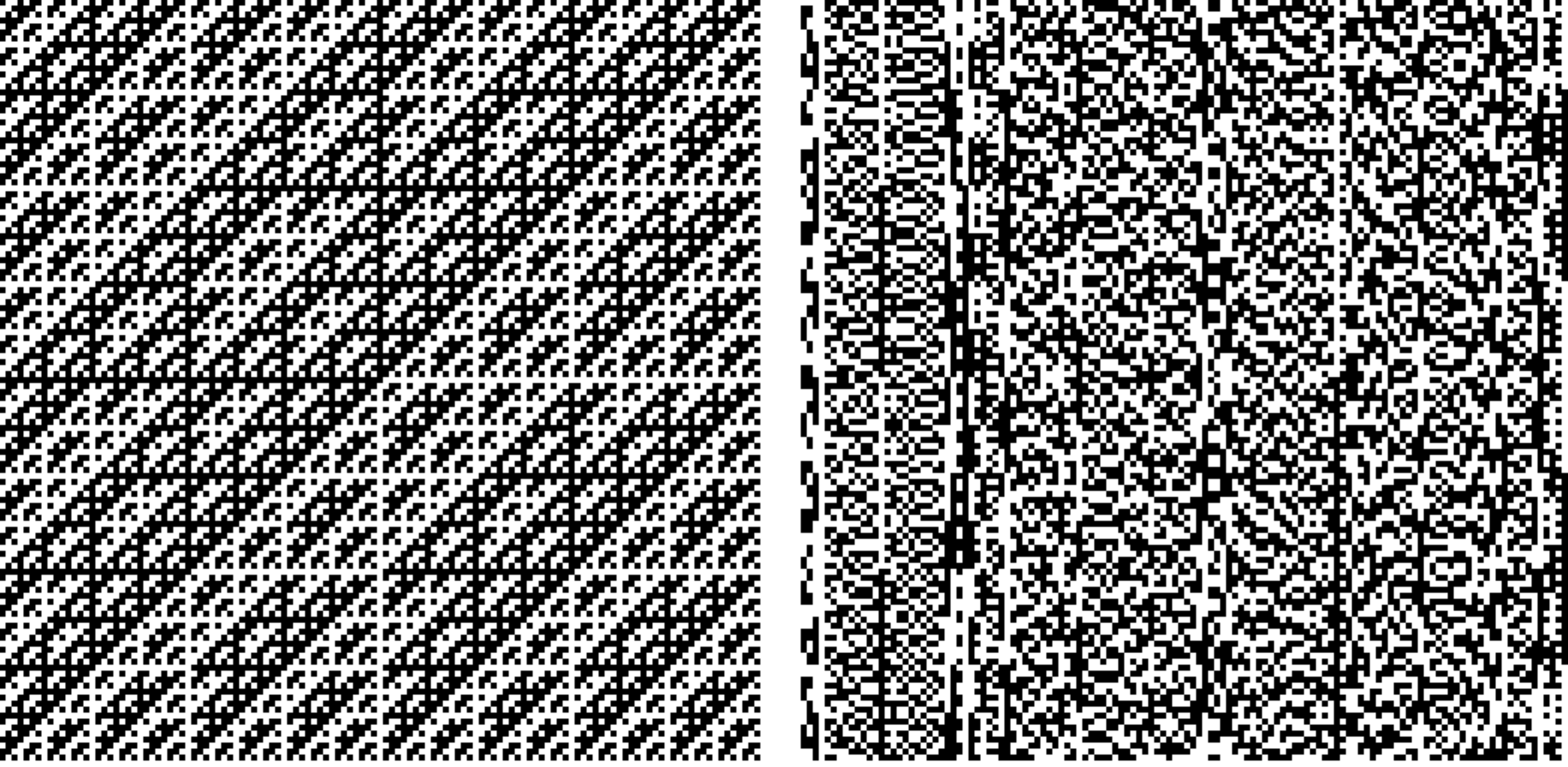}
	\caption{(Left) Hadamard pattern with a sampling resolution of 128\,$\times$\,128 pixels. (Right) Scrambled Hadamard pattern with a sampling resolution of 128\,$\times$\,128 pixels.}
	\label{fig:hadamard}
\end{figure}

The original image displayed on the DMD has fairly intricate details and the reconstructed image on the right hand side manages to reproduce some, but not all, of them. The colour rendition is accurate and the main artefact left to correct in the reconstructed image is the presence of the vertical lines at different locations across the picture.  These vertical lines appear in the same positions in other captured data, which suggests that they are not caused by random noise but more likely by the reconstruction process.

\subsection{Pattern Choice}\label{SecPattern}
\begin{figure}
	\centering\includegraphics[width=8cm]{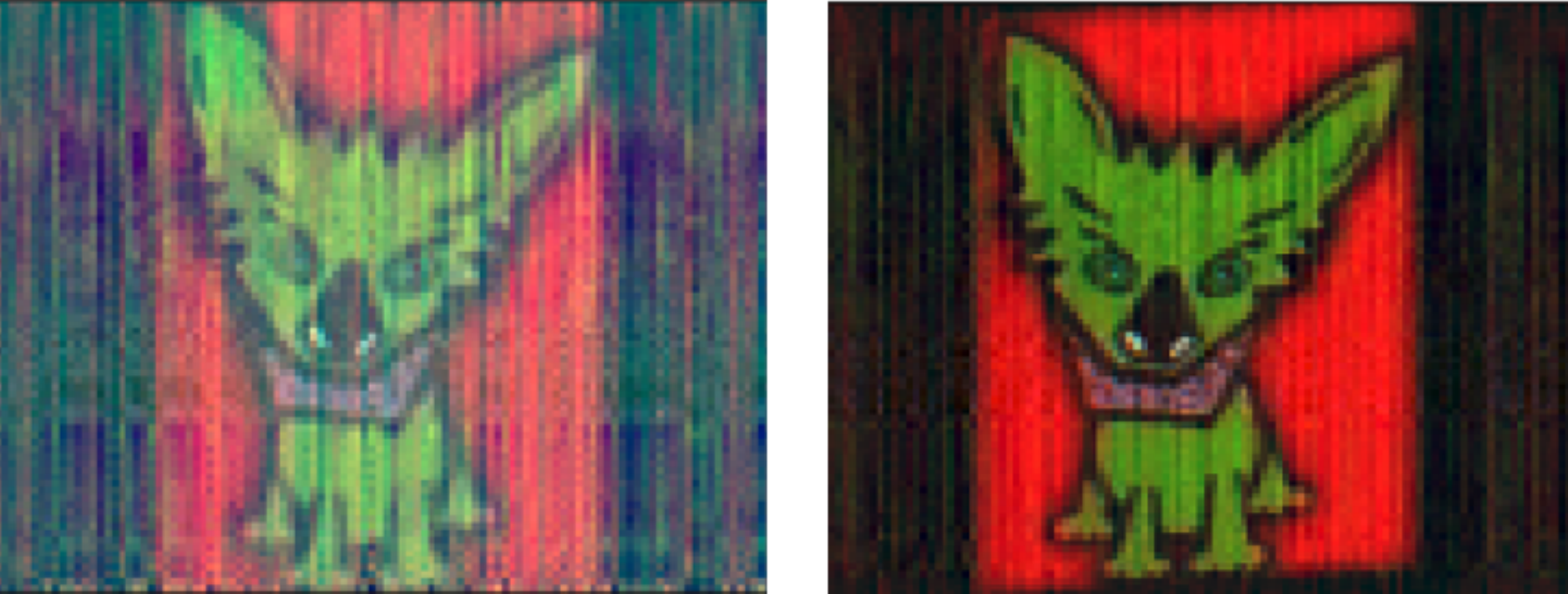}
	\caption{(Left) Reconstructed image using the scrambled Hadamard pattern with the DMD plane parallel to the LCD monitor ($\theta$=\ang{0}). (Right) Reconstructed image using the scrambled Hadamard pattern with the DMD plane parallel to the camera sensor ($\theta$=\ang{24})}
	\label{fig:Scrambled}
\end{figure}

The presence of the vertical lines always located in the same positions led us to investigate the effects of different patterns on the reconstructed image. The choice of a new pattern follows from the observation that the white calibration values have a 2D structure which cannot be represented in a 1D vector. This is especially important in the middle of the pattern where the corresponding row consists of all black followed by all white Hadamard pattern pixels, hence why the dark lines always appear where the sequence of black and white pixels of the columns are more heavily unbalanced. The new chosen pattern is a scrambled version of the Hadamard pattern used previously, where the columns have been permuted to avoid long constant value sequences (see Figure~\ref{fig:hadamard}). The result from this new pattern can be seen on the left hand side of Figure~\ref{fig:Scrambled}. It is clear that the reconstructed image is now free of the highly recognisable dark vertical lines of Figure~\ref{fig:64}.

To improve the quality of the reconstructed image, we changed the orientation of the DMD to be parallel to the camera sensor (situation depicted in Figure~\ref{fig:setup}), i.e. $\theta$=\ang{24}, at the cost of the scene no longer appearing square --- a parallelogram is fine, so long as the motion is perpendicular to the 1D summation, as this can be corrected after reconstruction.  The result can be seen in Figure~\ref{fig:Scrambled}: the colour fidelity is greatly increased and the black areas of the reconstructed image are very dark while good brightness is maintained elsewhere.

\subsection{High Resolution Mask} \label{Highres}
\begin{figure}
	\centering\includegraphics[width=8cm]{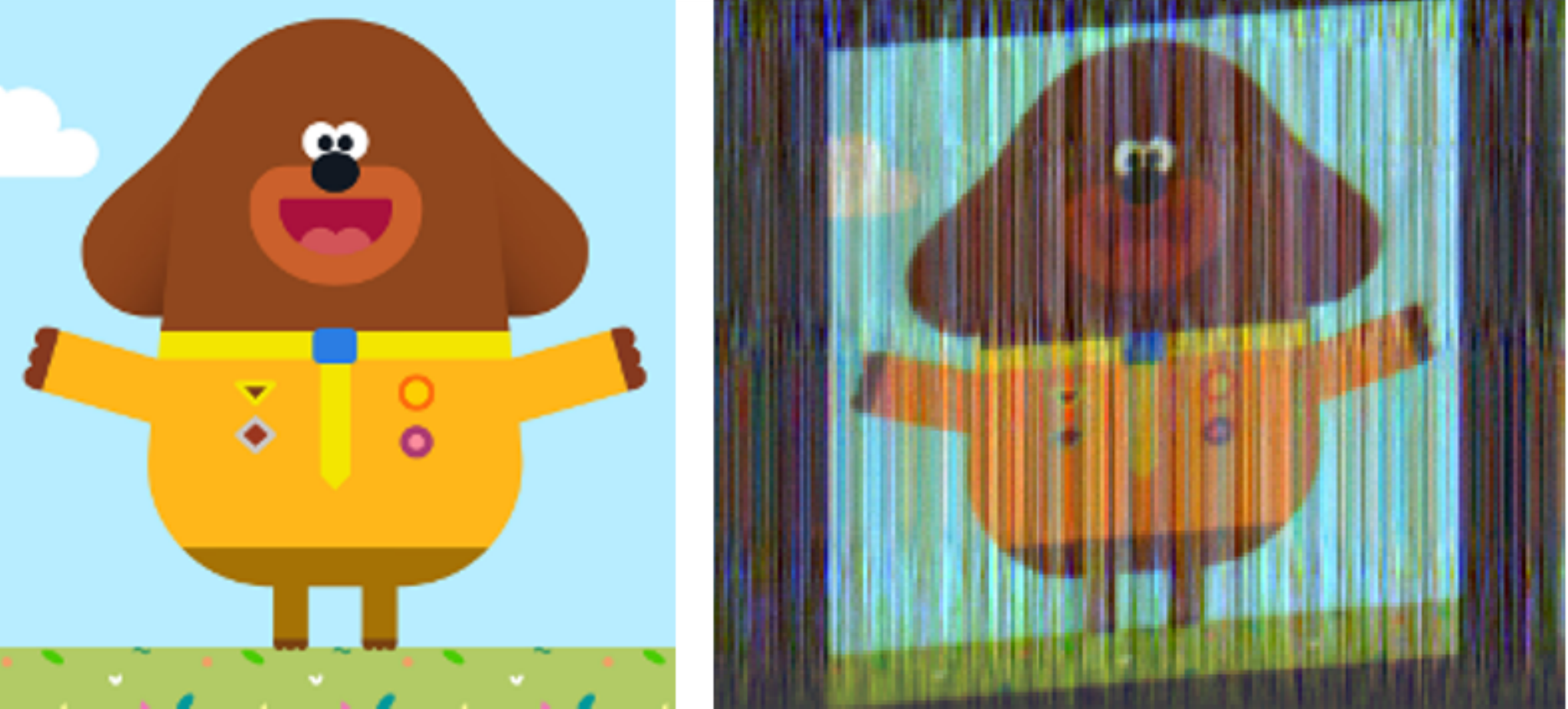}
	\caption{(Left) Image projected by the LCD monitor that is then moved across the DMD pattern. (Right) Reconstructed image of the object depicted on the left hand side, using the 256\,$\times$\,256 pixels scrambled Hadamard pattern.}
	\label{fig:256}
\end{figure}

The resolution of the reconstructed image is only limited by the resolution of the Hadamard pattern displayed on the DMD. The only thing to adjust when using a finer DMD pattern is the step size of the moving scene, which becomes smaller, requiring greater calibration accuracy to match with the Hadamard pattern on the DMD. A slight offset in the step size induces some noise in the reconstruction that cannot be corrected in post-processing. The difficulty here is that the scene becomes blurry on the edge of the DMD pattern because the latter is sitting at an angle ($\theta$=\ang{24}) with respect to the plane of the moving scene (see Section~\ref{Setup}), hindering straightforward step-size calibration methods. In order to take full advantage of the higher resolution mask of 256\,$\times$\,256 pixels we use a different scene with more details which should be more challenging to reconstruct. The scene as well as the reconstructed image can be seen in Figure~\ref{fig:256}. Unsurprisingly the reconstructed image is a lot more noisy than that of the 128\,$\times$\,128 pixels pattern, this is due to a combination of things. The more challenging nature of the scene, with finer details, makes it less forgiving to reconstruct. This, in combination with a small step size miscalibration of the moving scene, will lead to a significant increase in noise and will wash out most of the finer details of the scene.

The noise present in the reconstructed image led us to investigate the contribution of the different sources of noise to the overall degradation in image quality. We have identified three main noise contributors. Perhaps the most obvious one is the step size miscalibration of the moving scene, discussed above. This is purely hardware-limited and requires very careful calibration of the step size but can be almost entirely eliminated after precise calibration. The second source of noise comes from the lack of resolution the optics provide compared to that of the camera sensor, this is directly related to the sharpness of the image on the sensor and can only be improved by using better imaging optics. The camera used in the experiment has a resolution of 1280\,$\times$\,1024 pixels with a sensor diagonal of 1/1.8" and a pixel pitch of 5.3~$\mu$m. It means that each pixel of the DMD pattern (256\,$\times$\,256 pixels) is ideally imaged on to sixteen pixels on the camera sensor, which corresponds to a surface area of about 20\,$\times$\,20~$\mu$m$^2$. This is well below the performance of the lens used in our experimental setup, which is estimated to be between three to four times larger according to our ray tracing software. This could be reduced by using high-end imaging optics or custom-made optics tailored to match the performance required for our application. The lack of resolution also affects the contrast of the DMD pattern image on the camera. This is particularly problematic when the dark pixels of the pattern have a non zero value and contribute directly to the weightings used in reconstruction (see Section~\ref{Reconstruction}). Similarly, the contrast could also be improved by using better imaging optics as well as minimising stray light entering the system reaching the DMD or the imaging sensor. Lastly, due to the 1D light integration scheme of a pushframe camera (as explained in Section~\ref{SecPattern} it is only possible to correct the uniformity in one dimension) it is important to achieve very good uniformity across the DMD pattern.

\begin{figure}
	\centering\includegraphics[width=8cm]{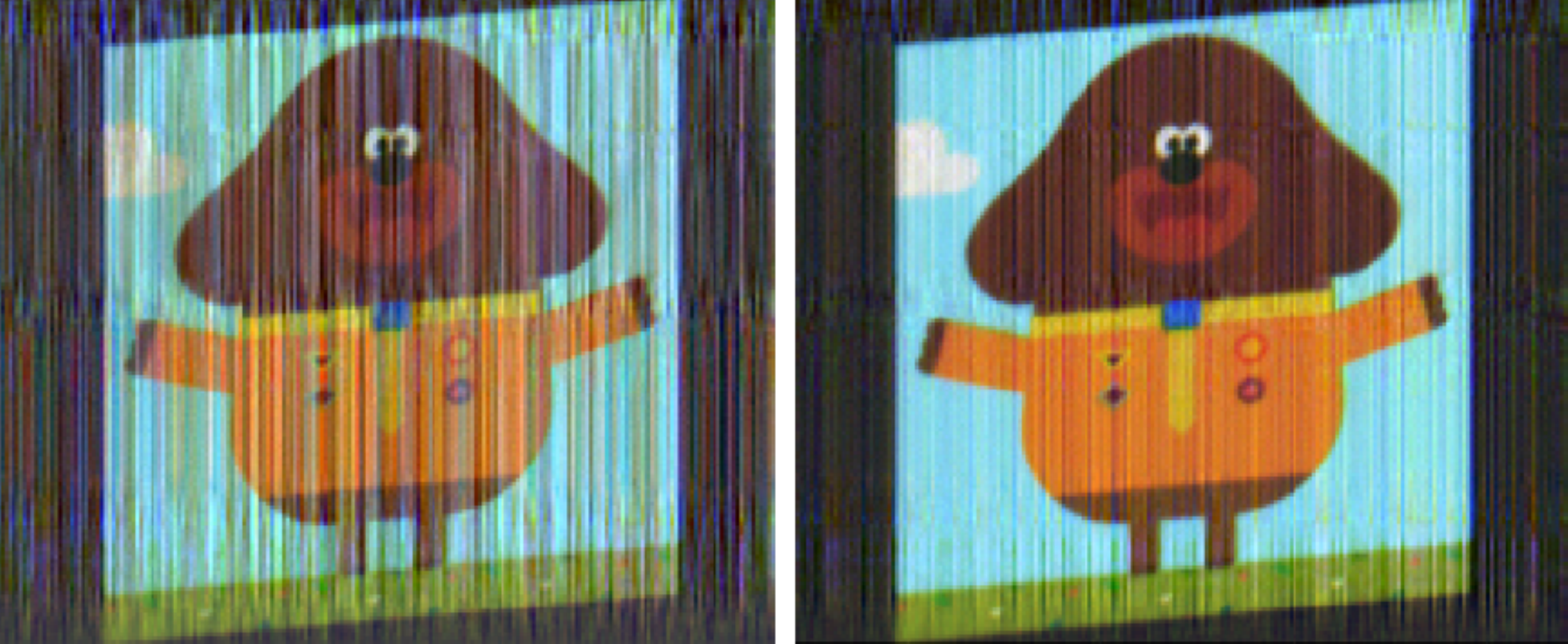}
	\caption{(Left) Reconstructed image using the 256\,$\times$\,256 pixels scrambled Hadamard pattern with 1D uniformity correction. (Right) Same reconstructed image using 2D uniformity and contrast correction.}
	\label{fig:noise}
\end{figure}

The pushframe camera scheme introduced in this paper involves the presence of a 1D light integration system as explained in Section~\ref{Setup}. However, we decided to perform the one dimensional light integration in post-processing rather than using custom-made hardware for demonstration purposes. This allows us to use correction algorithms that we can apply to the two dimensional image captured by the camera. Indeed, it is possible to apply a 2D illumination uniformity compensation to the image of the DMD pattern, thus making it perfectly uniform. The 2D image also allows us to correct for infinite contrast by setting the value of the black pattern pixels to zero. The result after applying these corrections is shown in Figure~\ref{fig:noise}. The image is a lot cleaner and most of the noise is actually suppressed from the reconstructed image. This is an encouraging result which indicates that if good uniformity were achieved as well as  using better imaging optics, we would be able to improve the quality of the reconstructed image. It is important to note that it would not have been possible to apply these two corrections if the 1D integration was done via hardware. This configuration allows us to understand the limitations of our experimental setup and pinpoint the areas of improvement necessary to improve the quality of the reconstructed image. Unfortunately, the significant amount of noise in the uncorrected reconstructed image means there is nothing to be gained by increasing the resolution further.

\section{Discussion}
The current setup is capable of producing images with good fidelity as seen in Figure~\ref{fig:Scrambled} and also achieved promising results at higher resolutions. However, the reasons for the drop in image quality at the highest resolution tested (256\,$\times$\,256 pixels) with our setup can be easily identified, as explained in Section~\ref{Highres}. The main limitations are hardware driven, especially the imaging optics, which if replaced could improve the quality of the reconstructed images as observed in Figure~\ref{fig:noise}. The performance of a simple achromatic doublet lens used in the experimental setup cannot match that of multi-element lenses commonly used in photography in order to exploit the ever decreasing pixel size and ever increasing resolution of emerging multi-megapixel sensors.

The prohibitive cost of high performance custom-made optics (necessary for the 1D optical compression realisation) makes it less than ideal to commit to for an experimental setup that could not be taken out of a laboratory. It could, however, be a more viable route for a prototype that we would then be able to showcase and test in different real world environments but it is outside the scope of this paper. Instead, the next natural step would be to buy an off-the-shelf magnifying imaging lens that can resolve the resolution of our current camera sensor. Along with better optics, it would also be useful to use a bigger sensor. The camera used in the experiment has a sensor with a diagonal of 1/1.8" and 1" sensors that fit in the same mechanical housing are available. That would effectively double the size of the sensor and help decrease the strain on the optical performance of the imaging optics. Another way of improving the quality of the reconstructed image would be to use an object mounted on a high precision translation stage that would give us precise control over the alignment of the moving scene with the DMD pattern.

In this paper we showed that it is possible to create a pushframe camera with a visible wavelength (colour) sensor and perform the 1D optical integration in post-processing. However, this restricts the number of spectral bands the camera can resolve to those provided by the camera (red, green and blue), as well as having the bandwidth of each wavelength set by the performance of the Bayer filter located on top of the sensor. This situation is not ideal and it would be beneficial to have more than three spectral bands in the visible in order to make this pushframe camera a true multispectral tool for remote sensing. 

In order to achieve multiple bands in the visible range it is necessary to perform the 1D optical integration using custom-made optics. The idea would be to compress the image from the DMD down to a few hundred microns on the camera sensor before placing a diffractive element in front of the latter, thus providing multiple bands to be analysed. The number of bands that the camera can resolve would depend on the performance of the compression optics: the higher the compression the easier it is to separate the different bands. Another solution would be to compress the image down to 0.5~mm, i.e. the maximum pixel height of commercially-available linear arrays in the visible range, and use multiple linear arrays stacked on top of each other where the separation between each array would determine the wavelength to be observed on the different linear sensors. Alternatively, we could also separate each wavelength individually using a set of different bandpass filters and then separately perform the optical compression for each linear array. The latter solution has the disadvantage of requiring more than one set of costly 1D integration optics but gives more freedom in both the wavelength selection and bandwidth due to the availability of bandpass filters in the visible. We propose that a system using a diffraction grating along with a single large area sensor would meet both the space and weight requirements for nano-satellite deployment and represent a viable alternative to pushbroom and snapshot cameras for remote sensing.

We have just seen that there is a clear path forward for hardware improvements, however there is also room for software improvements. In this paper we have only used one kind of pattern, a Hadamard matrix, that works very well for reconstruction but is not a pattern of choice for compressive sampling. For remote sensing, on a nano-satellite, power consumption and therefore data exchange between Earth and the space platform needs to be optimised. It is therefore worthwhile to implement a suitable pattern in an instrument designed to be mounted in a spacecraft. The working principle of a pushframe camera, i.e. the scene moving across the different columns of the DMD pattern, could potentially pave the way to an adaptive measurement scheme. Indeed, one could think of an algorithm that would be able to detect certain objects, such as a ship, after crossing only a small portion of the DMD pattern, and then change the remaining pattern columns to an optimised pattern tailored for the object previously detected. Similarly, it would be advantageous to implement an adaptive resolution technique that would allow the pattern to modify its resolution based on the scene under observation. For instance if the camera was observing the ocean, when the scene was very uniform, the pattern would switch to a lower resolution because there is nothing of interest to observe. Now imagine a ship in the vast ocean background: the software would pick up on the new object in the scene (before it is outside the field of view of the camera) and would adapt both its resolution and pattern to prioritise the subsequent reconstruction of the newly detected ocean vessel. This powerful combination would allow the spacecraft to only transmit data when something of significant importance has been detected. A pushframe camera would also be fertile ground for machine learning where a camera could be trained for the detection of specific targets.

\section{Conclusion}
In this paper, we have made the first experimental demonstration of a 1D integrating pushframe camera. We have shown that the system is capable of producing good quality images at various resolutions. We have demonstrated that the new approach on the single pixel camera scheme presented in this article has the potential to fill the gap between a pushbroom camera and its more conventional snapshot counterpart. This new camera becomes particularly relevant for space applications, low light environments where the high SNR would be beneficial and remote sensing where data exchange needs to be minimised. The compactness of the setup makes it an ideal candidate for space deployment in a constellation of nano-satellites, each specialised in the detection of different objects. The constellation of spacecraft could be tailored to suit the needs of the client by using specific patterns and algorithms to detect and perform the reconstruction of very specific objects. The main advantage here, is that it is possible to change the pattern remotely which means that the constellation of satellites could be rapidly retasked. We believe the technology to be very promising and that the compactness, versatility and performance of such a camera would make it an ideal candidate for remote sensing.

\section*{Funding}
Funded by the UK Space Agency through NSTP3-PF-031 and the CEOI-11 Call project ‘Compact Multispectral Imager for Nanosatellites II‘. 


\section*{Acknowledgments}
The authors thank 'Thorlabs, Inc' for using their 'Face Value Dog Artwork' and 'Studio AKA' for using 'Duggee'.

\section*{Disclosures}

The authors declare no conflicts of interest.


\bibliography{mybib}{}

\begin{thebibliography}{10}

\bibitem{abdo2018dual}
Mohammad Abdo, Erik F{\"o}rster, Patrick Bohnert, Vlad Badilita, Robert
  Brunner, Ulrike Wallrabe, and Jan~G Korvink.
\newblock Dual-mode pushbroom hyperspectral imaging using active system
  components and feed-forward compensation.
\newblock {\em Review of Scientific Instruments}, 89(8):083113, 2018.

\bibitem{adao2017hyperspectral}
Telmo Ad{\~a}o, Jon{\'a}{\v{s}} Hru{\v{s}}ka, Lu{\'\i}s P{\'a}dua, Jos{\'e}
  Bessa, Emanuel Peres, Raul Morais, and Joaquim Sousa.
\newblock Hyperspectral imaging: A review on {UAV}-based sensors, data
  processing and applications for agriculture and forestry.
\newblock {\em Remote Sensing}, 9(11):1110, 2017.

\bibitem{anderson2009challenges}
JA~Anderson and MS~Robinson.
\newblock Challenges utilizing pushframe camera images.
\newblock In {\em Lunar and Planetary Science Conference}, volume~40, 2009.

\bibitem{arablouei2016fast}
Reza Arablouei, Ethan Goan, Stephen Gensemer, and Branislav Kusy.
\newblock Fast and robust pushbroom hyperspectral imaging via {DMD}-based
  scanning.
\newblock In {\em Novel Optical Systems Design and Optimization XIX}, volume
  9948, page 99480A. International Society for Optics and Photonics, 2016.

\bibitem{arnob2018compressed}
Md~Masud~Parvez Arnob, Hung Nguyen, Zhu Han, and Wei-Chuan Shih.
\newblock Compressed sensing hyperspectral imaging in the 0.9--2.5 $\mu$m
  shortwave infrared wavelength range using a digital micromirror device and
  {InGaAs} linear array detector.
\newblock {\em Applied optics}, 57(18):5019--5024, 2018.

\bibitem{baraniuk2006method}
Richard Baraniuk, Dror Baron, Marco Duarte, Ilan Goodman, Don Johnson, Kevin
  Kelly, Courtney Lane, Jason Laska, Dharmpal Takhar, Michael Wakin, et~al.
\newblock Method and apparatus for compressive imaging device, October~26 2006.
\newblock US Patent App. 11/379,688.

\bibitem{candes2005decoding}
EJ~Candes and T~Tao.
\newblock Decoding by linear programming.
\newblock {\em IEEE Transactions on Information Theory}, 51(12):4203--4215,
  2005.

\bibitem{candes2006near}
EJ~Candes and T~Tao.
\newblock Near-optimal signal recovery from random projections: Universal
  encoding strategies?
\newblock {\em IEEE Transactions on Information Theory}, 12(52):5406--5425,
  2006.

\bibitem{deutsch2009adaptive}
Shay Deutsch, Amir Averbush, and Shay Dekel.
\newblock Adaptive compressed image sensing based on wavelet modeling and
  direct sampling.
\newblock In {\em SAMPTA'09}, 2009.

\bibitem{duarte2008single}
Marco~F Duarte, Mark~A Davenport, Dharmpal Takhar, Jason~N Laska, Ting Sun,
  Kevin~F Kelly, and Richard~G Baraniuk.
\newblock Single-pixel imaging via compressive sampling.
\newblock {\em IEEE signal processing magazine}, 25(2):83--91, 2008.

\bibitem{edgar2015simultaneous}
Matthew~P Edgar, Graham~M Gibson, Richard~W Bowman, Baoqing Sun, Neal Radwell,
  Kevin~J Mitchell, Stephen~S Welsh, and Miles~J Padgett.
\newblock Simultaneous real-time visible and infrared video with single-pixel
  detectors.
\newblock {\em Scientific Reports}, 5:10669, 2015.

\bibitem{fowler2014compressive}
James~E Fowler.
\newblock Compressive pushbroom and whiskbroom sensing for hyperspectral
  remote-sensing imaging.
\newblock In {\em 2014 IEEE International Conference on Image Processing
  (ICIP)}, pages 684--688. IEEE, 2014.

\bibitem{gibson2017real}
Graham~M Gibson, Baoqing Sun, Matthew~P Edgar, David~B Phillips, Nils Hempler,
  Gareth~T Maker, Graeme~PA Malcolm, and Miles~J Padgett.
\newblock Real-time imaging of methane gas leaks using a single-pixel camera.
\newblock {\em Optics Express}, 25(4):2998--3005, 2017.

\bibitem{gowen2007hyperspectral}
AA~Gowen, CPo O'Donnell, PJ~Cullen, G~Downey, and JM~Frias.
\newblock Hyperspectral imaging--an emerging process analytical tool for food
  quality and safety control.
\newblock {\em Trends in Food Science \& Technology}, 18(12):590--598, 2007.

\bibitem{henriksson2017imaging}
Markus Henriksson.
\newblock An imaging system parallelizing compressive sensing imaging,
  September~28 2017.
\newblock US Patent App. 15/504,939.

\bibitem{kyriakides2016target}
Ioannis Kyriakides.
\newblock Target tracking using adaptive compressive sensing and processing.
\newblock {\em Signal Processing}, 127:44--55, 2016.

\bibitem{lochocki2016single}
Benjamin Lochocki, Adrian Gamb{\'\i}n, Silvestre Manzanera, Esther Irles,
  Enrique Tajahuerce, Jesus Lancis, and Pablo Artal.
\newblock Single pixel camera ophthalmoscope.
\newblock {\em Optica}, 3(10):1056--1059, 2016.

\bibitem{lu2014medical}
Guolan Lu and Baowei Fei.
\newblock Medical hyperspectral imaging: a review.
\newblock {\em Journal of Biomedical Optics}, 19(1):010901, 2014.

\bibitem{musarra2019non}
Gabriella Musarra, Ashley Lyons, E~Conca, Yoann Altmann, Federica Villa,
  F~Zappa, Miles~J Padgett, and Daniele Faccio.
\newblock Non-line-of-sight three-dimensional imaging with a single-pixel
  camera.
\newblock {\em Physical Review Applied}, 12(1):011002, 2019.

\bibitem{phillips2017adaptive}
David~B Phillips, Ming-Jie Sun, Jonathan~M Taylor, Matthew~P Edgar, Stephen~M
  Barnett, Graham~M Gibson, and Miles~J Padgett.
\newblock Adaptive foveated single-pixel imaging with dynamic supersampling.
\newblock {\em Science Advances}, 3(4):e1601782, 2017.

\bibitem{polak2019use}
Adam Polak, Fraser~K Coutts, Paul Murray, and Stephen Marshall.
\newblock Use of hyperspectral imaging for cake moisture and hardness
  prediction.
\newblock {\em IET Image Processing}, 13(7):1152--1160, 2019.

\bibitem{qiao2015design}
Yang Qiao, Xiping Xu, Tao Liu, and Yue Pan.
\newblock Design of a high-numerical-aperture digital micromirror device camera
  with high dynamic range.
\newblock {\em Applied Optics}, 54(1):60--70, 2015.

\bibitem{radwell2014single}
Neal Radwell, Kevin~J Mitchell, Graham~M Gibson, Matthew~P Edgar, Richard
  Bowman, and Miles~J Padgett.
\newblock Single-pixel infrared and visible microscope.
\newblock {\em Optica}, 1(5):285--289, 2014.

\bibitem{rani2018systematic}
Meenu Rani, SB~Dhok, and RB~Deshmukh.
\newblock A systematic review of compressive sensing: Concepts, implementations
  and applications.
\newblock {\em IEEE Access}, 6:4875--4894, 2018.

\bibitem{shapiro2008computational}
Jeffrey~H Shapiro.
\newblock Computational ghost imaging.
\newblock {\em Physical Review A}, 78(6):061802, 2008.

\bibitem{stantchev2016noninvasive}
Rayko~Ivanov Stantchev, Baoqing Sun, Sam~M Hornett, Peter~A Hobson, Graham~M
  Gibson, Miles~J Padgett, and Euan Hendry.
\newblock Noninvasive, near-field terahertz imaging of hidden objects using a
  single-pixel detector.
\newblock {\em Science Advances}, 2(6):e1600190, 2016.

\bibitem{starck2010sparse}
Jean-Luc Starck, Fionn Murtagh, and Jalal~M Fadili.
\newblock {\em Sparse image and signal processing: wavelets, curvelets,
  morphological diversity}.
\newblock Cambridge University Press, 2010.

\bibitem{sun2012normalized}
Baoqing Sun, Stephen~S Welsh, Matthew~P Edgar, Jeffrey~H Shapiro, and Miles~J
  Padgett.
\newblock Normalized ghost imaging.
\newblock {\em Optics Express}, 20(15):16892--16901, 2012.

\bibitem{sun2016single}
Ming-Jie Sun, Matthew~P Edgar, Graham~M Gibson, Baoqing Sun, Neal Radwell,
  Robert Lamb, and Miles~J Padgett.
\newblock Single-pixel three-dimensional imaging with time-based depth
  resolution.
\newblock {\em Nature Communications}, 7:12010, 2016.

\bibitem{tajahuerce2014image}
Enrique Tajahuerce, Vicente Dur{\'a}n, Pere Clemente, Esther Irles, Fernando
  Soldevila, Pedro Andr{\'e}s, and Jes{\'u}s Lancis.
\newblock Image transmission through dynamic scattering media by single-pixel
  photodetection.
\newblock {\em Optics Express}, 22(14):16945--16955, 2014.

\bibitem{van2012multi}
Freek~D Van~der Meer, Harald~MA Van~der Werff, Frank~JA Van~Ruitenbeek, Chris~A
  Hecker, Wim~H Bakker, Marleen~F Noomen, Mark Van Der~Meijde, E~John~M
  Carranza, J~Boudewijn De~Smeth, and Tsehaie Woldai.
\newblock Multi-and hyperspectral geologic remote sensing: A review.
\newblock {\em International Journal of Applied Earth Observation and
  Geoinformation}, 14(1):112--128, 2012.

\end{thebibliography}
\bibliographystyle{plain}






\end{document}